\documentclass[11pt]{article}
\usepackage[left=1in,top=1in,right=1in,bottom=1in]{geometry} % Does NOT work right now
\usepackage{times}
\usepackage{booktabs}
\usepackage{threeparttable}

\makeatletter
\renewcommand{\paragraph}{%
  \@startsection{paragraph}{4}%
  {\z@}{1ex \@plus 1ex \@minus .2ex}{-1em}%
  {\normalfont\normalsize\bfseries}%
}
\makeatother

\usepackage{tikz}
\usetikzlibrary{backgrounds,positioning,fit,patterns,shadows,calc}

\usepackage{epsfig}
\usepackage{amsfonts}
\usepackage{amssymb}
\usepackage{amstext}
\usepackage{amsmath}

\usepackage{calligra}
\usepackage[T1]{fontenc}

\usepackage{amsthm, thmtools}

\usepackage{paralist}

\usepackage{nameref}
\definecolor{ForestGreen}{rgb}{0.1333,0.5451,0.1333}
\definecolor{DarkRed}{rgb}{0.8,0,0}
\usepackage[linktocpage=true,
	pagebackref=true,colorlinks,
	linkcolor=DarkRed,citecolor=ForestGreen,
	bookmarks,bookmarksopen,bookmarksnumbered]
	{hyperref}
\usepackage{cleveref}

\usepackage{thm-restate} % See section 1.4 of the pdf above

\usepackage{xspace}
\usepackage{color}
\usepackage{enumitem}
\usepackage{comment}
\usepackage{caption}
\usepackage{subcaption}
\usepackage[noend]{algorithmic}
\usepackage[section,boxed]{algorithm}

\makeatletter
\def\thmt@refnamewithcomma #1#2#3,#4,#5\@nil{%
  \@xa\def\csname\thmt@envname #1utorefname\endcsname{#3}%
  \ifcsname #2refname\endcsname
    \csname #2refname\expandafter\endcsname\expandafter{\thmt@envname}{#3}{#4}%
  \fi
}
\makeatother

\declaretheorem[numberwithin=section,refname={Theorem,Theorems},Refname={Theorem,Theorems}]{theorem}
\declaretheorem[numberlike=theorem,refname={Lemma,Lemmas},Refname={Lemma,Lemmas}]{lemma}

\renewcommand{\varepsilon}{\epsilon}

\newcommand{\cut}{
\ifmmode \text{{\calligra C}\ } \else{\calligra C}\xspace\fi
}

\newcommand{\id}{\ensuremath{{\sf id}}\xspace}

\def\poly{\operatorname{poly}}

\newboolean{short}
\setboolean{short}{false}

\newcommand{\shortOnly}[1]{\ifthenelse{\boolean{short}}{#1}{}}
\newcommand{\onlyShort}[1]{\ifthenelse{\boolean{short}}{#1}{}}
\newcommand{\longOnly}[1]{\ifthenelse{\boolean{short}}{}{#1}}
\newcommand{\onlyLong}[1]{\ifthenelse{\boolean{short}}{}{#1}}

\onlyShort{

\setlength{\textheight}{9.2in}
\setlength{\textwidth}{6.55in}
}

\newcommand{\squishlist}{
 \begin{list}{$\bullet$}
  { \setlength{\itemsep}{0pt}
     \setlength{\parsep}{2pt}
     \setlength{\topsep}{2pt}
     \setlength{\partopsep}{0pt}
     \setlength{\leftmargin}{1.5em}
     \setlength{\labelwidth}{1em}
     \setlength{\labelsep}{0.5em} } }
\newcommand{\squishend}{
  \end{list}  }

\ifdefined\ShowComment

\def\danupon#1{\marginpar{$\leftarrow$\fbox{D}}\footnote{$\Rightarrow$~{\sf #1 --Danupon}}}

\else

\def\danupon#1{}

\fi

\begin{document}
%\begin{titlepage}
\title{Brief Announcement: Almost-Tight Approximation\\ Distributed Algorithm for Minimum Cut} 
\author{
Danupon Nanongkai\thanks{ICERM, Brown University, USA. \hbox{E-mail}:~{\tt danupon@gmail.com}.}  
}

\date{}

%\pagenumbering{roman}
\maketitle 

\begin{abstract}
In this short paper, we present an improved algorithm for approximating the minimum cut on distributed (CONGEST) networks. Let $\lambda$ be the minimum cut. Our algorithm can compute $\lambda$ exactly in $\tilde O((\sqrt{n}+D)\poly(\lambda))$ time, where $n$ is the number of nodes (processors) in the network, $D$ is the network diameter, and $\tilde O$ hides $\poly\log n$. By a standard reduction, we can convert this algorithm into a $(1+\epsilon)$-approximation  $\tilde O((\sqrt{n}+D)/\poly(\epsilon))$-time algorithm. The latter result improves over the previous $(2+\epsilon)$-approximation $\tilde O((\sqrt{n}+D)/\poly(\epsilon))$-time algorithm of Ghaffari and Kuhn [DISC 2013]. Due to the lower bound of $\tilde \Omega(\sqrt{n}+D)$ by Das Sarma et al. [SICOMP 2013], this running time is {\em tight} up to a $\poly\log n$ factor. Our algorithm is an extremely simple combination of Thorup's tree packing theorem [Combinatorica 2007],  Kutten and Peleg's tree partitioning algorithm [J. Algorithms 1998], and Karger's dynamic programming [JACM 2000].
\end{abstract}

%\footnote{Ghaffari and Kuhn also provided a $(O(D)+\tilde O(n^{1/2+\epsilon}))$-time $O(\epsilon^{-1})$-approximation algorithm. This algorithm gives a slightly better running time  when $D$ is large ($O(D)$ instead of $\tilde O(D)$), but with a worse approximation guarantee.}

%\begin{center}
%{\bf Brief Announcement Submission}
%\end{center}

%\newpage
%\thispagestyle{empty}
%\tableofcontents

%\newpage
%\listoftheorems
%\end{titlepage}

\section{Introduction}

%Computing network connectivity, be it the minimum node or edge cut, is a fundamental algorithmic graph problem with many important network applications. 

%One of these applications is relevant to transferring information between nodes of a network, which is the ultimate goal of communication networks and also a central issue in distributed computing (e.g. \cite{PelegBook}).

%Finding or approximating minimum cuts are fundamental algorithmic graph problems with many important applications. One of these applications is relevant to transferring information between nodes of a network, which is the ultimate goal of communication networks and also a central issue in distributed computing (e.g. \cite{PelegBook}).

\paragraph{Problem.} In this paper, we study the time complexity of the fundamental problem of computing {\em minimum cut} on distributed network. Given a graph $G$, edge weight assignment $w$, and any set $X$ of nodes in $G$, the cut $\cut(X)$ is defined as 
$$\cut(X)=\sum_{(x, y)\in E(G), x\in X, y\notin X} w(x, y).$$ 
Our goal is to find $\lambda(G)=\min_{\emptyset \neq X\subsetneq V} \cut(X)$. 

\paragraph{Communication Model.} We use a standard message passing network (the CONGEST model \cite{PelegBook}). Throughout the paper, we let $n$ be the number of nodes and $D$ be the diameter of the network. Every node is assumed to have a unique ID, and initially knows the weights of edges incident to it. 
The execution in this network proceeds in synchronous rounds and in each round, each node can send a message of size $O(\log n)$ bits to each of its neighbors.  The goal of the problem is find the minimum or approximately minimum cut $X$. (Every node outputs whether it is in $X$ in the end of the process.) The time complexity is the number of rounds needed to compute this. (For more detail, see \cite{GhaffariK13}.)

%We consider this problem on the standard message-passing distributed networks in the CONGEST model \cite{PelegBook}. Due to space limitation, we refer to, e.g. \cite{PelegBook,GhaffariK13,DasSarmaHKKNPPW11}

\paragraph{Previous Work.} The current best algorithm is by Ghaffari and Kuhn \cite{GhaffariK13} which takes $\tilde O(\sqrt{n}+D)$ time with an approximation ratio of $(2+\epsilon)$. ($\tilde O(\cdot)$ hides the $O(\poly\log n)$ factor.) The running time of this algorithm matches the lower bound of Das Sarma et al. \cite{DasSarmaHKKNPPW11} which showed that this problem cannot be computed faster than $\tilde \Omega(\sqrt{n}+D)$ even when we allow a large approximation ratio. (This lower bound was also shown to hold even when a quantum communication is allowed \cite{ElkinKNP12}, and when a capacity of an edge is proportional to its weight \cite{GhaffariK13}.) For a more comprehensive literature review, see \cite{GhaffariK13}. 
%
%(building on \cite{PelegR00,Elkin06,KorKP13}) 
%Moreover, \cite{Censor-HillelGK13}

\paragraph{Our Results.} Our main result is a distributed algorithm that can compute the minimum cut {\em exactly}  in $\tilde O((\sqrt{n}+D)\poly(\lambda))$ time. For the case where the minimum cut is small (i.e. $\tilde O(1)$), the running time of our algorithm matches the lower bound \cite{DasSarmaHKKNPPW11,GhaffariK13}. When the minimum cut is large, Karger's edge sampling technique \cite{Karger94} can be used to reduce the minimum cut to $\tilde O(1)$ with the cost of $(1+\epsilon)$ approximation factor (due to the space limit, we refer the readers to \cite[Lemma 7]{Thorup07} for the statement of Karger's sampling result). This makes our algorithm a $(1+\epsilon)$-approximation $\tilde O(\sqrt{n}+D)$-time one, improving the previous algorithm of Ghaffari and Kuhn \cite{GhaffariK13}.

\paragraph{Techniques.} Our algorithm is a simple combination of techniques from \cite{Thorup07,KuttenP98,Karger00}. The starting point of our algorithm is Thorup's tree packing theorem, which shows that if we generate $\Theta(\lambda^7 \log^3 n)$ trees $T_1, T_2, \ldots$, where tree $T_i$ is the minimum spanning tree with respect to the loads induced by  $\{T_1, \ldots, T_{i-1}\}$, then one of these trees will contain exactly one edge in the minimum cut. (Due to the space limit, we refer the readers to \cite[Theorem~9]{Thorup07} for the full statement.) Since we can use the $\tilde O(\sqrt{n}+D)$-time algorithm of Kutten and Peleg \cite{KuttenP98} to compute the minimum spanning tree (MST), the problem of finding a minimum cut is reduced to finding the minimum cut that {\em $1$-respects a tree}; i.e., finding which edge in a given spanning tree defines a smallest cut (see the formal definition in \Cref{sec:one respect tree}). Solving this problem is our main result. 

To solve this problem, we usea simple observation of Karger \cite{Karger00} which reduces the problem to computing the sum of degree and the number of edges contained in a subtree rooted at each node. 
%Karger \cite{Karger00} use this observation to develop a dynamic programming to solve this problem in the centralized setting. 
We use this observation along with Kutten and Peleg's {\em tree partitioning} \cite{KuttenP98} to quickly compute these quantities. This requires several (elementary) steps, which we will discuss in more detail in \Cref{sec:one respect tree}. 

\paragraph{Concurrent Result.} Independent from our work, Su \cite{Su14} also achieved a $(1+\epsilon)$-approximation $\tilde O(\sqrt{n}+D)$-time algorithm for this problem. His starting point is, like ours, Thorup's theorem \cite{Thorup07}. The way he finds the minimum cut that $1$-respects a tree is, however, very different. In particular, he uses edge sampling to make the minimum cut of a certain graph be one and use Thurimella's algorithm \cite{Thurimella97} to find a bridge. (See Algorithm~2 in \cite{Su14} for details.) This gives a nice and simple way to achieve essentially the same approximation result as ours, with a small drawback that minimum cut cannot be computed exactly, even when it is small. 

%The time complexity of this problem has been widely open until very recently. 

%Our slight advantage is that we can compute small cut exactly. 

\newcommand{\figscale}{0.3\textwidth}
\begin{figure}
\centering
\begin{subfigure}{\figscale}
\centering
\includegraphics[page=1, clip=true, trim=7cm 9.9cm 8cm 1.1cm, width=\textwidth]{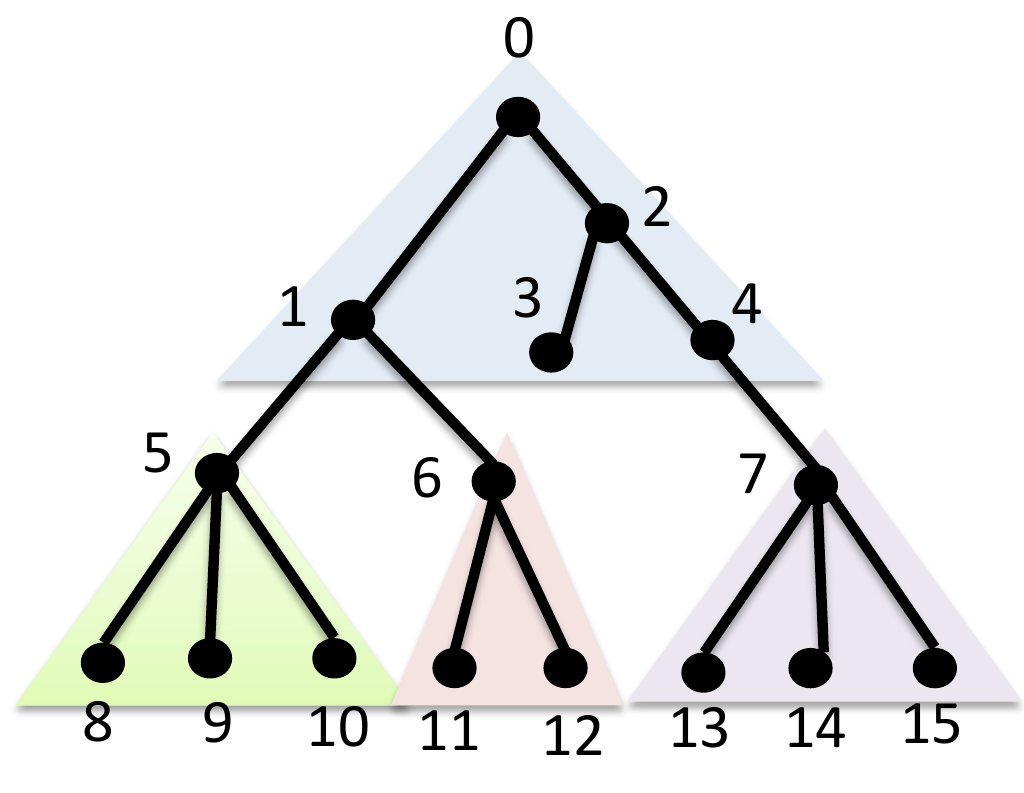}
\caption{}\label{fig:one}
\end{subfigure}
\hspace{0.01\textwidth}
\begin{subfigure}{\figscale}
\centering
\includegraphics[page=2, clip=true, trim=7cm 9.9cm 8cm 1.1cm, width=\textwidth]{mincut.pdf}
\caption{}\label{fig:two}
\end{subfigure}
\hspace{0.01\textwidth}
\begin{subfigure}{\figscale}
\centering
\includegraphics[page=3, clip=true, trim=7cm 9.9cm 8cm 1.1cm, width=\textwidth]{mincut.pdf}
\caption{}\label{fig:three}
\end{subfigure}

\vspace{0.3cm}

\begin{subfigure}{\figscale}
\centering
\includegraphics[page=4, clip=true, trim=7cm 9.9cm 8cm 1.1cm, width=\textwidth]{mincut.pdf}
\caption{}\label{fig:four}
\end{subfigure}
\hspace{0.01\textwidth}
\begin{subfigure}{\figscale}
\centering
\includegraphics[page=5, clip=true, trim=7cm 9.9cm 8cm 1.1cm, width=\textwidth]{mincut.pdf}
\caption{}\label{fig:five}
\end{subfigure}
\hspace{0.01\textwidth}
\begin{subfigure}{\figscale}
\centering
\includegraphics[page=6, clip=true, trim=7cm 9.9cm 8cm 1.1cm, width=\textwidth]{mincut.pdf}
\caption{}\label{fig:six}
\end{subfigure}
\caption{}\label{fig:example}
%\\
%(\subref{fig:one}) Graph $G$ with tree $T$ (marked by bold (black) edges) rooted at node of ID $0$ and fragments (defined by triangular regions) obtained from \Cref{subsec:fragment}. \\%
%(\subref{fig:two}) Tree $T_F$ (nodes in this tree are the triangles) representing the structure between fragments. This tree is known to every node after we finish the steps in \Cref{subsec:inter fragment tree}.\\ 
%(\subref{fig:three}) Tree $T'_F$ representing fragments and their least common ancestors. This tree is known to every node after we finish the steps in\Cref{subsec:LCA of fragments}.}\label{fig:example}
\end{figure}

\section{Distributed Algorithm for Finding a Cut that 1-Respects a Tree}\label{sec:one respect tree}

In this section, we solve the following problem: Given a spanning tree $T$ on a network $G$ rooted at some node $r$, we want to find an edge in $T$ such that when we cut it, the cut define by edges connecting the two connected component of $T$ is smallest. To be precise, for any node $v$, define $v^\downarrow$  to be the set of nodes that are descendants of $v$ in $T$, including $v$.
\danupon{To do: Define $\cut(X)$.}
%
%\begin{definition}[$v^\downarrow$ and $v^\uparrow$]
%For any node $v$ define $v^\downarrow$ to be the set of nodes that are descendants of $v$ in $T$, including $v$. Similarly, define $v^\uparrow$ to be the set of nodes that are ancestors of $v$ in $T$, including $v$. 
%\end{definition}
%
The problem is then to compute $c^* = \min_{v\in V(G)} \cut(v^\downarrow)$.
%
%\begin{align}
%c^* = \min_{v\in V(G)} \cut(v^\downarrow).\label{eq:min one tree cut}
%\end{align}
%
%consider any edge $(u, v)$. Let $T_1$ and $T_2$ be the two connected components resulting from cutting $(u, v)$ in $T$. Define $cut_T(u, v)$ be the number of edges between $T_1$ and $T_2$ in $G$. Then, we want to compute $\min_{(u, v)\in E(T)} cut_T(u, v)$. 
%
%We show the following. 
%
\begin{theorem}[Main Result] %[Main Result of \Cref{sec:one respect tree}]
There is an $\tilde O(n^{1/2}+D)$-time distributed algorithm that can compute $c^*$ as well as find a node $v$ such that  $c^* = \cut(v^\downarrow)$. 
\end{theorem}
%
%While we state our result as computing the {\em value} of $c^*$, our algorithm will in fact find the cut  
%
%%We note that our algorithm in facts compute $\cut(v^\downarrow)$ for all $v$ ($v$ knows )
%
In fact, at the end of our algorithm every node $v$ knows $\cut(v^\downarrow)$. Our algorithm is inspired by the following observation used in Karger's dynamic programming \cite{Karger00}. For any node $v$, let $\delta(v)$ be the weighted degree of $v$, i.e. $\delta(v)=\sum_{u\in V(G)} w(u, v)$. Let $\rho(v)$ denote the total weight of edges whose endpoints' least common ancestor is $v$. Let $\delta^\downarrow(v)=\sum_{u\in v^\downarrow} \delta(u)$ and $\rho^\downarrow(v)=\sum_{u\in v^\downarrow} \rho(u)$. 

%It relies on the following simple lemma (Lemma 5.9 in \cite{Karger00}). 
%
%, which says that when we count the degree of all nodes, we will get the value of $\cut(v^\downarrow)$ 
%
%\begin{definition}[$\delta$ and $\rho$]
%For any node $v$, let $\delta(v)$ be the weighted degree of $v$, i.e. $\delta(v)=\sum_{u\in V(G)} w(u, v)$. Let $\rho(v)$ denote the total weight of edges whose endpoints' least common ancestor is $v$. Let $\delta^\downarrow(v)=\sum_{u\in v^\downarrow} \delta(u)$ and $\rho^\downarrow(v)=\sum_{u\in v^\downarrow} \rho(u)$. 
%\end{definition}

\begin{lemma}[Karger \cite{Karger00} (Lemma 5.9)]\label{thm:Karger}
$\cut(v^\downarrow) = \delta^\downarrow(v)-2\rho^\downarrow(v)$.
\end{lemma}

Our algorithm will make sure that every node $v$ knows $\delta^\downarrow(v)$ and $\rho^\downarrow(v)$. By \Cref{thm:Karger}, this will be sufficient for every node $v$ to compute $c^*$. 
%$\cut(v^\downarrow)$ as in \Cref{eq:min one tree cut}. 
The algorithm is divided in several steps, as folows. 

%... The problem reduces to computing $\delta(u)$ 

%\newcounter{step}
%\addtocounter{step}{1}
%\paragraph{Step \arabic{step}: Partition $T$} 

\danupon{TO DO: Define Diameter}

\paragraph{Step 1: Partition $T$ into Fragments and Compute ``Fragment Tree'' $T_F$.} %\label{subsec:fragment}
We use the algorithm of Kutten and Peleg \cite[Section 3.2]{KuttenP98} to partition nodes in tree $T$ into $O(\sqrt{n})$ subtrees, where each subtree has $O(\sqrt{n})$ diameter\footnote{To be precise, we compute a {\em $(\sqrt{n}+1, O(\sqrt{n}))$ spanning forest}. Also note that we in fact do not need this algorithm since we obtain $T$ by using Kutten and Peleg's MST algorithm, which already computes the $(\sqrt{n}+1, O(\sqrt{n}))$ spanning forest as a subroutine. See \cite{KuttenP98} for details.} 
(every node knows which edges incident to it are in the subtree containing it). This algorithm takes $O(n^{1/2}\log^*n+D)$ time. We call these subtrees {\em fragments} and denote them by $F_1, \ldots, F_k$, where $k=O(\sqrt{n})$. 
For any $i$, let $\id(F_i)=\min_{u\in F_i} \id(u)$ be the {\em ID of $F_i$}. We can assume that every node in $F_i$ knows $\id(F_i)$. This can be achieved in $O(\sqrt{n})$ time by a communication within each fragment. 

Let $T_F$ be a rooted tree obtained by contracting nodes in the same fragment into one node. This naturally defines the child-parent relationship between fragments (e.g. the fragments labeled (5), (6), and (7) in \Cref{fig:two} are children of the fragment labeled (0)). 
Let the {\em root} of any fragment $F_i$, denoted by $r_i$, be the node in $F_i$ that is nearest to the root $r$ in $T$. 
We now make every node know $T_F$: Every ``inter-fragment'' edge, i.e. every edge $(u, v)$ such that $u$ and $v$ are in different fragments, either node $u$ or $v$ broadcasts this edge and the IDs of fragments containing $u$ and $v$ to the whole network. This step takes $O(\sqrt{n})$ time since there are $O(\sqrt{n})$ edges in $T$ that link between different fragments. 
Note that this process also makes every node knows the roots of all fragments since, for every inter-fragment edge $(u, v)$, every node knows the child-parent relationship between two fragments that contain $u$ and $v$. 

\paragraph{Step 2: Compute Fragments in Subtrees of Ancestors.}  For any node $v$ let $F(v)$ be the set of fragments $F_i\subseteq v^\downarrow$. For any node $v$ in any fragment $F_i$, let $A(v)$ be the set of ancestors of $v$ in $T$ that are in $F_i$ or the parent fragment of $F_i$ (also let $A(v)$ contain $v$). (For example, \Cref{fig:three} shows $A(15)$.)
The goal of this step is to make every node $v$ knows (i) $A(v)$ and (ii) $F(u)$ for all $u\in A(v)$. 
%
%%The goal of this step is to make every node $v$ in any fragment $F_i$ know $(i)$ 
%%
%%
%%$F(u)$, for $u=v$ and every ancestor $u$ of $v$ in the $F_i$ and in the parent fragment of $F_i$. 
%
%
%%the following: , (ii) its ancestors in the $F_i$ and in parent fragment of $F_i$, and (iii) fragments that are contained in $v^\downarrow$. 

First, we make every node $v$ know $F(v)$: for every fragment $F_i$ we aggregate from the leaves to the root of $F_i$ (i.e. upcast) the list of child fragments of $F_i$. This takes $O(\sqrt{n})$ time since there are $O(\sqrt{n})$ fragments to aggregate. In this process every node $v$ receives a list of child fragments of $F_i$ that are contained in $v^\downarrow$. It can then use $T_F$ to compute fragments that are descendants of these child fragments, and thus compute {\em all} fragments contained in $v^\downarrow$. 
Next, we make every node $v$ in every fragment $F_i$ know $A(v)$: every node $u$ sends a message containing its ID down the tree $T$ until this message reaches the leaves of the child fragments of $F_i$. Since each fragment has diameter $O(\sqrt{n})$, this process takes $O(\sqrt{n})$ time.\danupon{Should say that diameter and number of messages is $O(\sqrt{n})$ if possible.}
With some minor modifications, we can also make every node $v$ know  $F(u)$ for all $u\in A(v)$: Initially every node $u$ sends a message $(u, F')$, for every $F'\in F(u)$, to its children. Every node $u$ that receives a message $(u', F')$ from its parents sends this message further to its children {\em if $F'\notin F(u)$}. (A message $(u', F')$ that a node $u$ sends to its children should be interpreted as ``$u'$ is the lowest ancestor of $u$ such that $F'\in F(u')$''.)

\paragraph{Step 3: Compute $\delta^\downarrow(v)$.} For every fragment $F_i$, we let $\delta(F_i)=\sum_{v\in F_i} \delta(v)$. 
For every node $v$ in every fragment $F_i$, we will compute $\delta^\downarrow(v)$ by separately computing (i) $\sum_{u\in F_i\cap v^\downarrow} \delta(u)$ and (ii)  $\sum_{F_j\in F(v)} \delta(F_j)$. 
The first quantity can be computed in $O(\sqrt{n})$ time by computing the sum within $F_i$ (every node $v$ sends the sum $\sum_{u\in F_i\cap v^\downarrow} \delta(u)$ to its parent). 
To compute the second quantity, it suffices to make every node know $\delta(F_i)$ for all $i$ since every node $v$ already knows $F(v)$.  To do this, we make every root $r_i$ know $\delta(F_i)$ in $O(\sqrt{n})$ time by computing the sum of degree of nodes within each $F_i$. Then, we can make every node know $\delta(F_i)$ for all $i$ by letting $r_i$ broadcast $\delta(F_i)$ to the whole network.

%The first quantity can be already computed without communication since every node $v$ knows all fragments in $v^\downarrow$ as well as $\delta(F_i)$ for all fragments $F_i$. The second quantity can be computed by computing the sum within $F_i$ . 

%
%We can easily make every node knows $\delta(F_i)$ for every $F_i$ in $O(\sqrt{n})$ time: we make $r_i$ knows $\delta(F_i)$ in $O(\sqrt{n})$ time by computing the sum within each $F_i$ and then let $r_i$ broadcast $\delta(F_i)$ to the whole network. 

\paragraph{Step 4: Compute Merging Nodes and $T'_F$.}
We say that a node $v$ is a {\em merging node} if there are two distinct children $x$ and $y$ of $v$ such that both $x^\downarrow$ and $y^\downarrow$ contain some fragments (e.g. nodes $0$ and $1$ in \Cref{fig:one}). In other words, it is a point where two fragments ``merge''. 
%
%$v^\downarrow$ contains some distinct fragments $F_i$ and $F_j$ and there is no child $u$ of $v$ such that both $F_i$ and $F_j$ are contained in $u^\downarrow$ (e.g. nodes $0$ and $1$ in \Cref{fig:one,fig:four}). 
%
Let $T'_F$ be the following tree: Nodes in $T'_F$ are both roots of fragments ($r_i$'s) and merging nodes. The parent of each node $v$ in $T'_F$ is its lowest ancestor in $T$ that appears in $T'_F$ (see \Cref{fig:four} for an example). Note that every merging node has at least two children in $T'_F$. This shows that there are $O(\sqrt{n})$ merging nodes. 
%
%The goal of this step is to make every node know the list of its ancestors in $T$ that appear in $T'_F$.  
%
%are (i) merging nodes and (ii) roots of fragments. 
%
The goal of this step is to let every node know $T'_F$. 

First, note that every node $v$ can easily know whether it is a merging node or not in one round by checking, for each child $u$, whether $u^\downarrow$ contains any fragment (i.e. whether $F(u)=\emptyset$). The merging nodes then broadcast their IDs to the whole network. (This takes $O(\sqrt{n})$ time since there are $O(\sqrt{n})$ merging nodes.)
Note further that every node $v$ in $T'_F$ knows its parent in $T'_F$ because its parent in $T'_F$ is one of the ancestors in $A(v)$. 
%every node $v$ in fragment $F_i$ knows its ancestors in $F_i$ and parent fragment of $F_i$; its parent in $T'_F$ is one of these ancestors. 
%
So, we can make every node knows $T'_F$ in $O(\sqrt{n})$ rounds by letting  every node in $T'_F$ broadcast the edge between itself and its parent in $T'_F$ to the whole network.

\paragraph{Step 5: Compute $\rho^\downarrow(v)$.} We now count, for every node $v$, the number of edges whose  least common ancestor (LCA) of its end-nodes are $v$. 
For every edge $(x, y)$ in $G$, we claim that $x$ and $y$ can compute the LCA of $(x, y)$ by exchanging $O(\sqrt{n})$ messages through edge $(x, y)$. Let $z$ denote the LCA of $(x, y)$. Consider three cases (see \Cref{fig:five}). 
{\em Case 1:}
First, consider when $x$ and $y$ are in the same fragment, say $F_i$. In this case we know that $z$ must be in $F_i$. Since $x$ and $y$ have the lists of their ancestors in $F_i$, they can find $z$ by exchanging these lists. 
In the next two cases we assume that $x$ and $y$ are in different fragments, say $F_i$ and $F_j$, respectively. 
{\em Case 2:} $z$ is {\em not} in $F_i$ and $F_j$. In this case, $z$ is a merging node such that $z^\downarrow$ contains $F_i$ and $F_j$. Since both $x$ and $y$ knows $T'_F$ and their ancestors in $T'_F$, they can find $z$ by exchanging the list of their ancestors in $T'_F$. 
{\em Case 3:} $z$ is in $F_i$ (the case where $z$ is in $F_j$ can be handled in a similar way). In this case $z^\downarrow$ contains $F_j$. Since $x$ knows $F(x')$ for all its ancestors $x'$ in $F_i$, it can compute its lowest ancestor $x''$ such that $F(x'')$ contains $F_j$. Such ancestor is the LCA of $(x, y)$. 

Now we compute $\rho^\downarrow(v)$ for every node $v$ by splitting edges $(x, y)$ whose LCA is $v$ into two types (see \Cref{fig:six}): (i) those that $x$ and $y$ are in different fragments from $v$, and (ii) the rest. 
For (i), note that $v$ must be a merging node. In this case one of $x$ and $y$ creates a message $\langle v\rangle$. We then count the number of messages of the form $\langle v\rangle$ for every merging node $v$ by computing the sum along the breadth-first search tree of $G$. This takes $O(\sqrt{n}+D)$ time since there are $O(\sqrt{n})$ merging nodes. 
For (ii), the node among $x$ and $y$ that is in the same fragment as $v$ creates and keeps a message $\langle v\rangle$. Now every node $v$ in every fragment $F_i$ counts the number of messages of the form $\langle v\rangle$ in $v^\downarrow\cap F_i$ by computing the sum through the tree $F_i$. Note that, to do this, every node $u$ has to send the number of messages of the form $\langle v\rangle$ to its parent, for all $v$ that is an ancestor of $u$ in the same fragment. There are $O(\sqrt{n})$ such ancestors, so we can compute the number of messages of the form $\langle v\rangle$ for every node $v$ {\em concurrently} in $O(\sqrt{n})$ time (by pipelining). 

%This completes the step.
%This completes the computation of $\rho^\downarrow(v)$. 

%It follows that we can count the number of messages for every node $v$ in every fragment in $O(\sqrt{n})$.

%\section{Open Problems}
%
%Open: Sublinear exact algorithm, $s$-$t$ ..., node connectivity. 
%
%Open problems: Exact mincut, Fully connected network, $s$-$t$ cut (connected to maxflow), 2-cut the tree. 
%
%Our algorithm can be used to test the edge-connectivity of graph in $\poly(\lambda) XXX$ time. Before Thurimell? can only produce a certificate. 

\paragraph{Acknowledgment:} The author would like to thank Thatchaphol Saranurak for bringing Thorup's tree packing theorem \cite{Thorup07} to his attention.

  \let\oldthebibliography=\thebibliography
  \let\endoldthebibliography=\endthebibliography
  \renewenvironment{thebibliography}[1]{%
    \begin{oldthebibliography}{#1}%
      \setlength{\parskip}{0ex}%
      \setlength{\itemsep}{0ex}%
  }%
  {%
    \end{oldthebibliography}%
  }
{ %\small
  \onlyShort{  \newpage}
\bibliographystyle{alpha}
\bibliography{references}
}
%\onlyLong{

\end{document}